\documentclass[onecolumn,trackchanges,printer]{aa}

\usepackage{graphicx,epsfig,fancyhdr,psfig,rotating,amsmath,epsf,txfonts,natbib,epstopdf,multirow,natbib,twoopt}
\usepackage{color}
\usepackage{ulem}
\usepackage{soul}
\usepackage{hyperref}
\usepackage{lineno}
\makeatletter
\usepackage{subfigure}
\usepackage{tabularx}
\usepackage{appendix}
\usepackage{dcolumn}
\usepackage{threeparttable}
\usepackage{longtable}

\def \halpha {H$\alpha$}
\def \kms {{\rm km\;s$^{-1}$}}
\def \arcsec {$^{''}$}

\def \caii {Ca\,{\sc ii}}

\def \heii {He\,{\sc ii}}
\def \chanel1330  {{1330}~\AA}
\def \chanel171 {{171}~\AA}
\def \chanel193 {{193}~\AA}

\def\typei{Type-\uppercase\expandafter{\romannumeral1}}
\def \typeii{Type-\uppercase\expandafter{\romannumeral2}}
\def \paperi{Paper~\uppercase\expandafter{\romannumeral1}}
\graphicspath{{figs/}}
\makeatletter
\newcommand{\bibnote}[2]{\global\@namedef{#1note}{#2}}
\newcommand{\biblink}[2]{\global\@namedef{#1link}{#2}}
\makeatother

\bibpunct{(}{)}{;}{a}{}{,}
\makeatletter
 \newcommandtwoopt{\citeads}[3][][]{%
   \nonstopmode
   \href{http://adsabs.harvard.edu/abs/#3}%
        {\def\hyper@linkstart##1##2{}%
         \let\hyper@linkend\@empty\citealp[#1][#2]{#3}}
   \biblink{#3}{\href{http://adsabs.harvard.edu/abs/#3}{ADS}}%
   \errorstopmode}            
 \newcommandtwoopt{\citepads}[3][][]{%
   \nonstopmode
   \href{http://adsabs.harvard.edu/abs/#3}%
        {\def\hyper@linkstart##1##2{}%
         \let\hyper@linkend\@empty\citep[#1][#2]{#3}}
   \biblink{#3}{\href{http://adsabs.harvard.edu/abs/#3}{ADS}}%
   \errorstopmode}            
 \newcommandtwoopt{\citetads}[3][][]{%
   \nonstopmode
   \href{http://adsabs.harvard.edu/abs/#3}%
        {\def\hyper@linkstart##1##2{}%
         \let\hyper@linkend\@empty\citet[#1][#2]{#3}}
   \biblink{#3}{\href{http://adsabs.harvard.edu/abs/#3}{ADS}}%
   \errorstopmode}            
 \newcommandtwoopt{\citeyearads}[3][][]{%
   \nonstopmode
   \href{http://adsabs.harvard.edu/abs/#3}%
        {\def\hyper@linkstart##1##2{}%
         \let\hyper@linkend\@empty\citeyear[#1][#2]{#3}}
   \biblink{#3}{\href{http://adsabs.harvard.edu/abs/#3}{ADS}}%
   \errorstopmode}            
\makeatother

\graphicspath{{./figs/}}

\begin{document}
\title{Statistical properties of \halpha\ jets in the polar coronal hole and their implications in coronal activities}
\titlerunning{\halpha\ jets in the polar coronal hole}
\authorrunning{Qi Y., Huang Z., Xia L., et al.}
\author{Youqian Qi\inst{1}\and Zhenghua Huang\inst{1}\and Lidong Xia\inst{1}\and Hui Fu\inst{1}\and Mingzhe Guo\inst{1}\and Zhenyong Hou\inst{1,2}\and Weixin Liu\inst{1}\and Mingzhe Sun\inst{1}\and Dayang Liu\inst{1}}

\institute{Shandong Provincial Key Laboratory of Optical Astronomy and Solar-Terrestrial Environment, Institute of Space Sciences, Shandong University, Weihai, 264209 Shandong, China\\
\email{xld@sdu.edu.cn}
\and
{\it Now at} School of Earth and Space Sciences, Peking University, Beijing 100871, China}

\abstract{Dynamic features, such as chromospheric jets, transition region network jets, coronal plumes and coronal jets, are abundant in the network regions of the solar polar coronal holes.}
{We investigate the relationship between chromospheric jets and coronal activities (e.g., coronal plumes and jets).} {We analyze observations of a polar coronal hole including the filtergrams that were taken by the New Vacuum Solar Telescope (NVST) at the \halpha\,--0.6\,\AA\ to study the \halpha\ jets, and the Atmospheric Imaging Assembly (AIA) 171\,\AA\ images to follow the evolution of coronal activities.} {\halpha\ jets are persistent in the network regions, only some regions (denoted as R1--R5) are rooted with discernible coronal plumes. With an automated method, we identify and track 1\,320 \halpha\ jets in the network regions. We find that the average lifetime, height and ascending speed of the \halpha\ jets are 75.38 s, 2.67 Mm, 65.60 \kms, respectively. The \halpha\ jets rooted in R1--R5 are higher and faster than those in the others. We also find that propagating disturbances (PDs) in coronal plumes have a close connection with the \halpha\ jets. The speeds of 28 out of 29 \halpha\ jets associated with PDs are $\gtrsim$50 \kms. In a case of coronal jet, we find that the speeds of both the coronal jet and the \halpha\ jet\ are over 150 \kms, suggesting that both cool and hot jets can be coupled together.} {Based on our analyses, it is evident that more dynamic \halpha\ jets could release the energies to the corona, which might be the results of the development of Kelvin-Helmholtz instability (KHi) or small-scaled magnetic activities. We suggest that chromospheric jets, transition region network jets and ray-like features in the corona are coherent phenomena, and they are important tunnels for cycling energy and mass in the solar atmosphere.}

\maketitle
\date{Received  / Accepted }
\section{Introduction} \label{sec:intro}
The chromosphere is dynamic and highly-structured, which links the photosphere
and the corona.
In the photosphere, the magnetic flux tubes build the magnetic network, which expands upward and
appears as bright patches in the chromosphere, namely, chromospheric network.
Therefore, the network boundaries in the chromosphere spatially coincide with the magnetic network concentrations in the photosphere.
The chromospheric network boundaries are abundant with jet-like, fine-scale features,
which are referred as spicules
at the limb, mottles in the quiet sun and fibrils in active regions \citep[\textit {e.g.,}][\textit {etc.}]
{1968SoPh....3..367B,2007ApJ...660L.169R,2007ApJ...655..624D}.
Spicules have been studied for more than a century and are still one of the most popular topics at the present days \citep[see \textit{ e.g.,}][for extensive overviews]{1968SoPh....3..367B,2000SoPh..196...79S,2012SSRv..169..181T}

\par
Thanks to high spatio-temporal resolution observations,
\citet{2007PASJ...59S.655D} pointed out that there are two different types of spicules (~\typei~$\&$~\typeii~).
The typical characteristic of~\typei~spicules ( also known as traditional spicules ) is that they
move up and down with the ascending speeds in the range of $15 - 40$\,\kms .
The~\typei~spicules have lifetimes of $150 - 400$\,s and heights of $4 - 8$\,Mm, and they are believed to result from magnetoacoustic shocks.
In contrast,~\typeii~spicules rise with much higher speeds of $\sim100$\,\kms and disappear
rapidly with shorter lifetimes of $50 - 150$\,s. Type II spicules are suggested to be driven by magnetic
reconnections. \citet{2007Sci...318.1574D} found that
the \typeii~spicules also undergo sideways swaying motion which is signature of Alfv{\'e}n waves permeating
with enough energy to heat the corona and accelerate the solar wind.
With high-quality observations, the \typeii~spicules have been found to combine three
different types of motions: field-aligned flows, swaying motions and torsional motions,
of which swaying and torsional motions are considered as a sign that spicules propagate along their axis
with Alfv{\'e}n waves at high speeds ranging from 100 -- 300\,\kms.
With multi-wavelength observations, \typeii~spicules often appeared to heat to the transition region temperatures,
even up to the coronal temperatures~\citep[\textit{e.g.,}][]{2011Sci...331...55D,2014Sci...346..315D,2015ApJ...799L...3R}.
In the advanced 2.5D radiative MHD simulations,~\citet{2017Sci...356.1269M} reproduced many observed
properties of type II spicules with $\sim 100$\,\kms speeds and $\sim 10$\,Mm heights, and also revealed
that spicules could heat plasma to the corona by dissipation of current which is due to ambipolar diffusion.
The further numerical study showed that
the dissipation through ambipolar diffusion of electrical currents could also explain why the observed spicules could reach
a speed exceeding 100~\kms and heat to coronal temperatures~\citep{2017ApJ...849L...7D}.

\par
In active region plage regions, \citet{2006ApJ...647L..73H} and \citet{2007ApJ...655..624D}
analyzed \halpha\ observations and found that fibrils appear as short, dynamic and spicule-like features
which have lifetimes of $3 - 6$\,minutes and show quasi-periodicity.
These dynamic fibrils undergo ascent and descent motions with velocities of $10 - 35$\,\kms~at
the initial phase, which are similar to that of \typei~spicules. Moreover, the numerical
simulations demonstrated that these fibril-like features also show wave-behaviors.
That could be formed when flows and oscillations leak from magnetic flux concentrations
into chromosphere~\citep[\textit{e.g.,}][]{2006ApJ...647L..73H,2007ApJ...655..624D,
2007ApJ...666.1277H,2011ApJ...743..142H}.

\par
In quiet-Sun regions, mottles show as dark and spicule-like features against the disk in the wings of \halpha. Similar to fibrils, the motions of some mottles also follow a parabolic path and they can be driven by shock waves~\citep[\textit{e.g.,}][]{2007ApJ...660L.169R,2007ASPC..368...65D}.

\par
With the spectroscopic data, \citet{2008ApJ...679L.167L} firstly used \caii~$854.2$\,nm
line to investigate the rapid blueshifted excursions (RBEs) on disk, which are a sudden widening
of line profile in the blue wing without an associated redshift.
Combined with SJ images of \halpha, these RBEs show as narrow streaks emanating from
network~\citep{2009ApJ...705..272R} .
The RBEs show velocities of $15 - 20$\,\kms and a mean lifetime of 45\,s.
There are $\sim10^{5}$ RBEs on the surface of disk at any moment.
Recently,~\citet{2012ApJ...752..108S,2013ApJ...764..164S}  found that RBEs also undergo three kinds of motions like spicules.
Owing to the similarity, RBEs are interpreted as the on-disk counterparts of \typeii~spicules and suggested to result from magnetic reconnection~\citep{2009ApJ...705..272R}.

\par
Jet-like features are not only seen in the chromosphere, but also in the transition region and corona.
In the transition region, network jets are small-scale structures, having lifetimes of $20 - 80$ s and
speeds of $80 - 250$ \kms. A number of them appear as ``inverse-Y" shape, which should
be driven by magnetic reconnection~\citep{2014Sci...346A.315T}. In the corona, coronal
jets appear as transient collimated beams. More findings suggest that coronal jets
are driven by magnetic reconnection and may be the source of mass and energy
input to the upper solar atmosphere and the solar wind~\citep[see][and the references therein]{2016SSRv..201....1R}.
Plumes belong to another type of collimated structures, which are relatively large and stable.
~\citep[see][and the references therein]{2011A&ARv..19...35W} and might supply  plasma and energy for solar wind
~\citep[\textit{e.g.,}][]{2010ApJ...709L..88T,2011ApJ...736..130T, 2014ApJ...794..109F,2015ApJ...806..273L,2021SoPh..296...22H}.

\par
There are PDs in coronal plumes with 10\% - 20\% intensity variations, periods of 10 - 30 minutes and speeds of 75 - 170\,\kms~\citep[\textit{e.g.,}][]{1998ApJ...501L.217D,2011A&A...528L...4K}.
Furthermore, ~\citet{2011ApJ...736..130T} discovered that high-speed quasi-periodic outflows
could be found in both plumes and inter-plume regions, and their speeds are similar to that of PDs. These outflows were suggested to be responsible for the generation of PDs~\citep{2010A&A...510L...2M}.
By following the evolution of  spicules,~\citet{2015ApJ...809L..17J} and~\citet{2015ApJ...807...71P}
concluded that the spicules can trigger PDs.

\par
In the base of coronal plumes, \citet{2008ApJ...682L.137R} found that coronal jets could
contribute to rise and change brightness in the pre-existed plumes.
Given that those coronal jets are rooted in the chromospheric network and are mainly triggered by
magnetic reconnection, coronal plumes could also be powered by magnetic reconnection
~\citep[for an extensive overview, see][]{2016SSRv..201....1R}.
Furthermore,
~\citet{2014ApJ...787..118R} and~\citet{2018ApJ...868L..27P} found that a great number of jets and
transient bright points frequently occurred in the bases of coronal plumes and these features
could be the main energy source for coronal plumes.
In fact, the scenario that coronal plumes are driven by the magnetic reconnection between
the unipolarity magnetic features and nearby small-scale bipolar has been proposed much earlier~\citep[\textit{e.g.,}][]
{1995ApJ...452..457W,1997ApJ...484L..75W,2008SoPh..249...17W}.

\par
Since there are many kinds of jet-like features rooted in network regions, it is natural to ask whether there is any connection among them or not.
Aiming at this question, \citet[][hereafter \paperi]{2019SoPh..294...92Q} developed an automatic method to identify and track network jets.
They statistically analyzed a set of transition region network jets in the same network region, and they found that network jets in the root of coronal plumes have lifetime, height and speed averaging at 45.6\,s, 8.1\arcsec\ and 131\,\kms, significantly more dynamic than those in the region without discernible plume, which have longer lifetime of 50.2\,s, smaller height of 5.5\arcsec\ and lower speed of 89\,\kms\ in average. \paperi\ suggested that only more energetic network jets (likely with speeds more than 100\,\kms) can feed plasma to the corona.
In the present study, we further investigate the possible relationship between chromospheric jet-like features (\textit{e.g.,}~spicules, fibrils, mottles or RBEs) and coronal activities.
To achieve this, we analyzed data obtained with the ground-based 1-m NVST~\citep{2014RAA....14..705L,
2016NewA...49....8X} installed at the Fuxian Lake Solar Observing Stations operated by Yunnan Astronomical
Observatories in China and the space-borne AIA~\citep{2012SoPh..275...17L}
and the Helioseismic and Magnetic Imager\,\citep[HMI,][]{2012SoPh..275..229S} aboard the
Solar Dynamics Observatory\,\citep[SDO,][]{2012SoPh..275....3P}.

\par
The paper is organized as follows.
The observations and methodology are described in Sect.~\ref{sec:obser}.
The results are shown in Sect.~\ref{sec:resstat}.
The discussion and conclusions are given in Sect.\,\ref{sect:discussion} and in Sect.\,\ref{sect:conclusions}, respectively.

\section{Observations}\label{sec:obser}
The data set analyzed in this study was taken on 2018 September 15 targeting on a polar coronal hole.
The observations include filtergrams taken by NVST at the \halpha\,--0.6\,\AA\ with a bandpass of 0.25\,\AA, EUV images taken at 171\,\AA\ and 193\,\AA\ passbands by AIA and line-of-sight magnetograms by HMI.

\par
The NVST data were taken from 09:29\,UT to 10:03\,UT including a series of \halpha~images with a cadence of 6 s and a spatial scale of 0.136 arcsec/pixel.
The field-of-view covers a 144\arcsec~$\times$~144\arcsec\ region.
The stabilization of the \halpha\ image series is performed with a fast sub-pixel image registration
algorithm by the instrument team~\citep{2015RAA....15..569Y}.

\par
The AIA and HMI data were downloaded from JSOC.
The cadences of the AIA 171\,\AA~data and HMI line-of-sight magnetograms are 12\,s and 45\,s, respectively.
The spatial resolutions of the AIA and HMI data are 1.2\arcsec.
The AIA and HMI data are prepared with standard procedures provided by the instrument teams,
and the level 1.5 data are analyzed.

\par
The \halpha~images have been transposed and rotated clockwise by 90 degrees to fit into the
Helioprojective-Cartesian coordinate system as used in AIA and HMI data. Bad images in \halpha~data
have been manually removed and replaced by artificial ones obtained from interpolation of nearby good frames.
The network lanes that present as magnetic concentrations on HMI magnetograms and clusters of bright dots on \halpha~images are used to align the corresponding images.
We make use of AIA 1600\,\AA~as reference to align the \halpha~images and 171\,\AA~images. And we manually check the alignment between 1600\,\AA~and 171\,\AA~using referent features, despite images from different passbands of AIA have been aligned with each other by the data processing pipeline.

\par
\begin{figure*}[!ht]
\centering
\includegraphics[trim=0cm 3cm 0cm 4cm,clip,width=\textwidth]{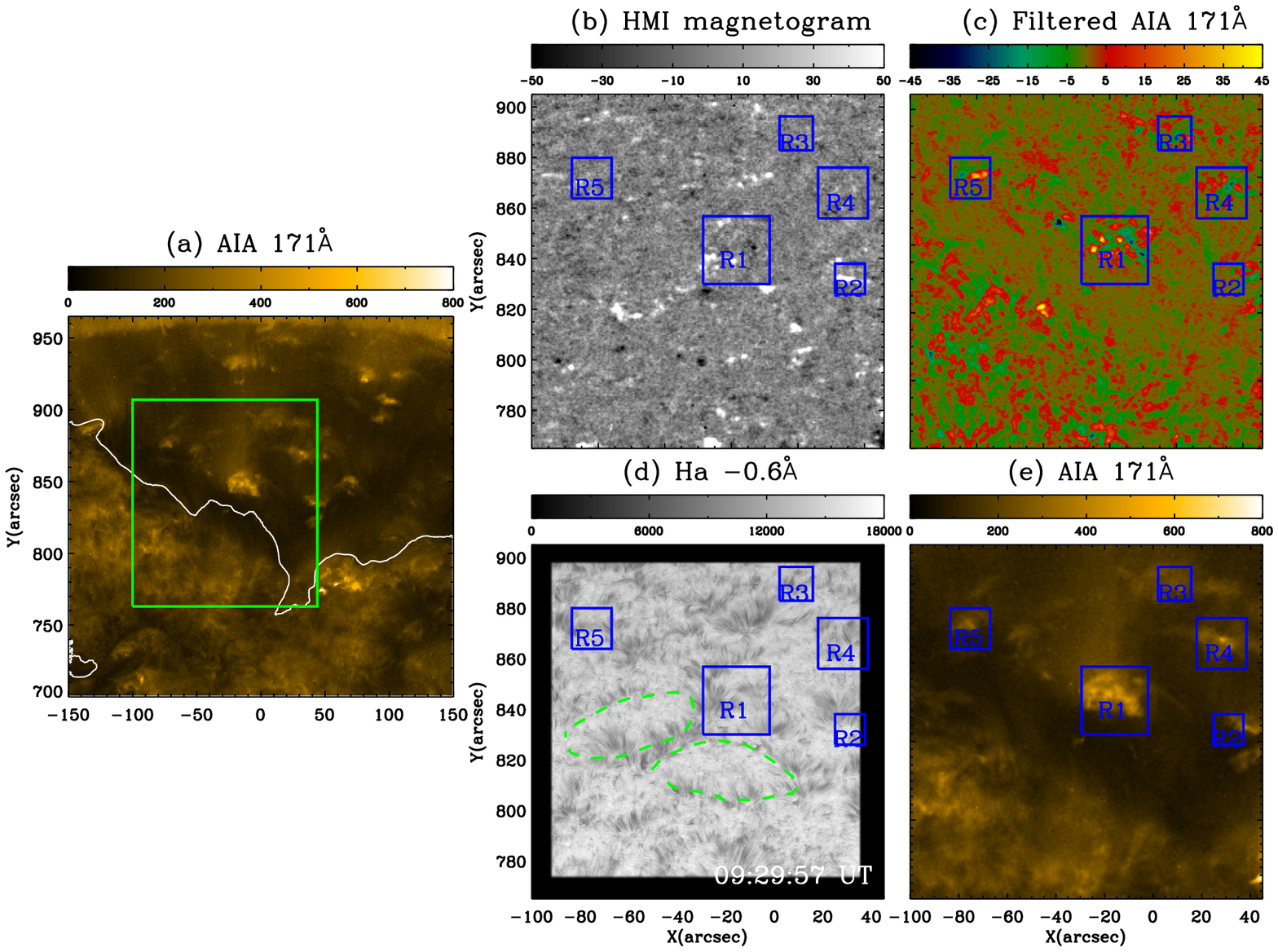}
\caption{Context images for regions studied in the present work taken on 2018 September 15. The AIA 171\,\AA\ image is giving an overview of the coronal structures in and around the studied regions (a). The contours (\textit{white lines}) outline the boundaries of the coronal holes determined from the AIA 193\,\AA\ image. The region
enclosed by the rectangle (green lines) is zoomed-in in panels (b--e). The region of interest seen with HMI magnetic features (b), AIA 171\,\AA\ filtered (10-30 minutes) image (c), NVST \halpha~$-0.6$\,\AA image (d) and AIA 171\,\AA\ image (e) is shown. The rectangles (\textit{blue lines}) in panels (b--e) mark the network regions where coronal plumes are clearly seen.
The green dotted lines in panel (d) mark two full and clear network regions.}
\label{figfov}
\end{figure*}

In Fig.\,\ref{figfov},
we show the studied regions seen with AIA, HMI and \halpha\,--0.6\,\AA~around 09:28\,UT.
The region is part of the polar coronal hole (see Fig.\,\ref{figfov}(a)).
In Fig.\,\ref{figfov}(b), even though the region is part of the polar region, we can see typical magnetic structures of network
regions~\citep[\textit{e.g.,}][]{2003A&A...399L...5X,2004A&A...424.1025X}. The network regions also
could be clearly seen in NVST \halpha\,--0.6\,\AA\ images (see the bright lanes
in Fig.\,\ref{figfov}(d)). We could clearly see lots of jet-like features rooting in the network lanes
in the \halpha\,--0.6\,\AA\ images. Given that this region is close to the north polar,
these jets exhibit upward motions from the \halpha~off-bands filter, so we defined
these jet-like features as the \halpha~jets in this paper.
In the coronal images (AIA 171\,\AA), we also see plume-like and/or jet-like features rooted in coronal bright points at some locations in the coronal hole (Figs.\,\ref{figfov}(a)\&(e)).
Figure\,\ref{figfov}(c) shows PDs with a simple bandpass filter method on the AIA 171\,\AA\ images.

\section{Data Analyses and Results}\label{sec:resstat}

In \halpha\ images, jets appear as dark features.
In Fig.~\ref{figfov} and the associated animation, we can see that \halpha~jets are abundant in the network regions.
To identify these jets, we upgrade the automatic algorithm that was described in \paperi\ and some key definitions are given in \citet{2017MNRAS.464.1753H}.
Note that
`` local peaks'' as defined in~\citet{2017MNRAS.464.1753H} are replaced by ``local trough'', or in other words, multiply the \halpha\ intensity by --1 to identify ``local peaks''.
In practical operations, we trace them from their bases to tops to obtain their full trajectories and then deduce their lifetimes and speeds.

\subsection{Statistics of \halpha\ jets}\label{subsec:alljets}
\begin{figure*}[!ht]
\centering
\includegraphics[trim=0cm 0cm 0cm 0cm,clip,width=\textwidth]{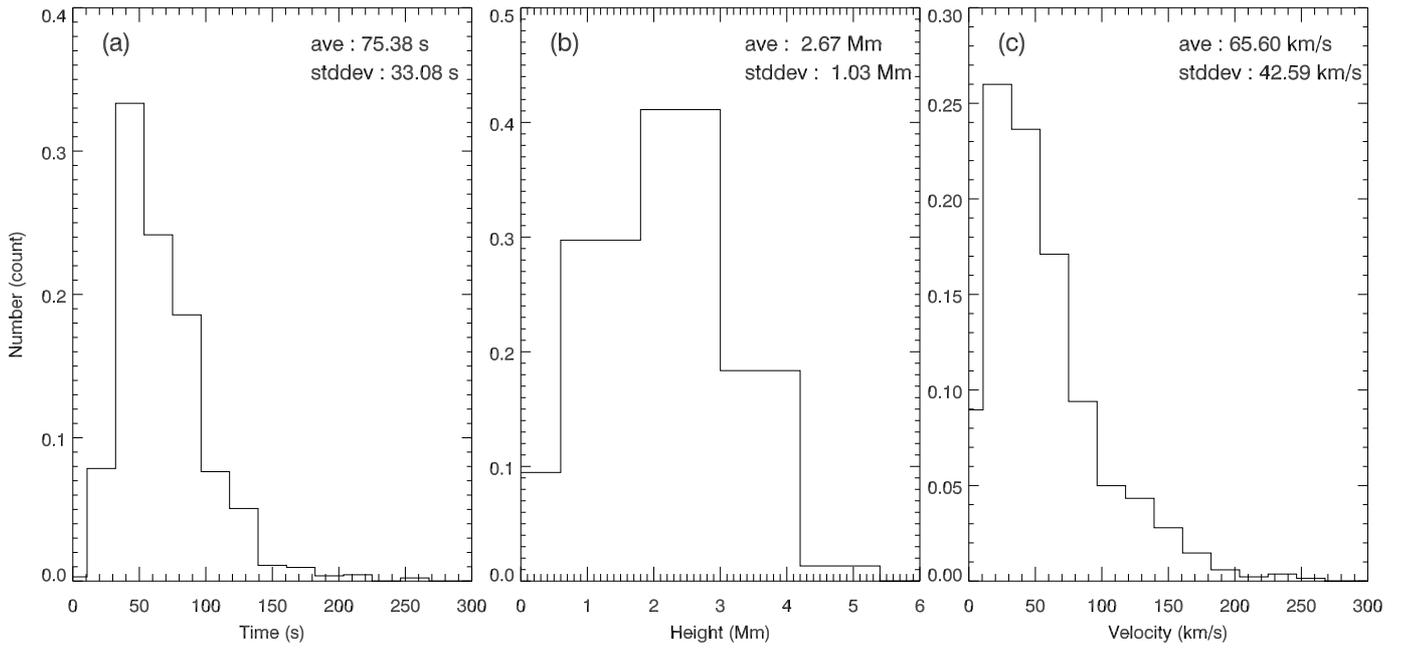}
\caption{The normalized distributions of the lifetimes (a), heights (b) and speeds (c) of the \halpha~jets identified and tracked in the whole field-of-view. The values following to``ave" and ``stddev" are the average values and standard deviations of the corresponding parameters, respectively.}
\label{figresult1}
\end{figure*}

By applying the automatic algorithm on the full field-of-view, we identify and trace 1\,320 \halpha\ jets during the observing period of time.
We proceed a statistics on their lifetimes, heights and speeds.
Figure~\ref{figresult1} shows the distributions of these properties.
We obtain an average lifetime of 75.38$\pm$33.08\,s,
an average height of 2.67$\pm$1.03\,Mm and an average
ascending velocity of 65.60$\pm$42.59\,\kms.
In the speed regime, 44\% of \halpha\ jets fall in the range between 10\,\kms ~and
50\,\kms.
These are consistent with those found by~\citet{2009SoPh..260...59P}, and this group of \halpha\ jets is likely triggered by shocks~\citep{1982ApJ...257..345H} and/or leakage of p-modes~\citep{2004Natur.430..536D}.
About 40\%\ of \halpha\ jets are in the range of 50--100\,\kms, which are in accord with~\typeii~spicules.
About 16\% of \halpha~jets have speeds higher than 100\,\kms, in which those having speeds more than 150\,\kms\ account for about 5\%. Such high velocities agree well with those reported for \typeii\ spicules off the limb~\citep[\textit{e.g.,}][]{2011Sci...331...55D,2017ApJ...849L...7D,2017Sci...356.1269M}. In observations, we also find plenty of \halpha~jets moving up to the highest point, then disappear.
We also find that quite a few \halpha~jets undergo obvious swaying motions, but it is not easy to
measure their periods because the lifetime of jets is too short to follow a complete cycle.
Lacking of the spectroscopic observations, the torsional motions could be not identified.

\par
\begin{figure*}[!ht]
\centering
\includegraphics[trim=0cm 0cm 0cm 0cm,clip,width=\textwidth]{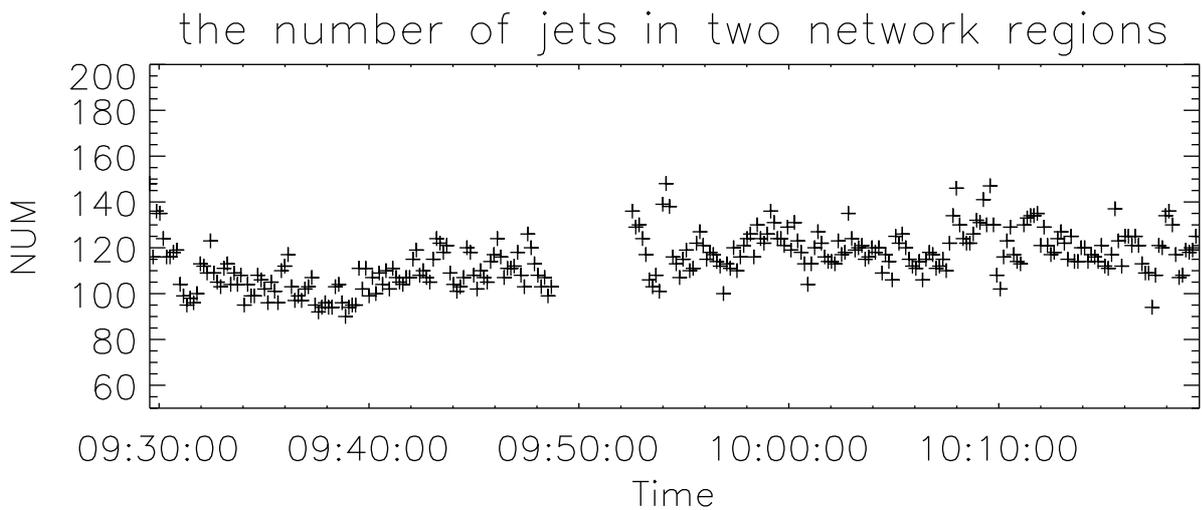}
\caption{Variation of the number of the \halpha~jets identified in two network regions as marked in Figure\,\ref{figfov}d.The data around 09:50:00 have been removed due to bad observing conditions.}
\label{birthrate}
\end{figure*}

From Fig.\,\ref{figfov}(d), there are numerous \halpha~jets in the network regions.
We select two full and clear network regions (see green-dashed lines) in Fig.\,\ref{figfov}(d) and count the
number of \halpha~jets as shown in Fig.\,\ref{birthrate}. The number of \halpha~jets is
$\sim 58$ per network region at any given time, which is consistent with reported spicules in
previous studies~\citep[\textit{e.g.,}][]{1973SoPh...30...63L,2011ApJ...731L..18M}.
Based on the similarity in lifetime, speed, height, spatial extent, location near network, birthrate,
we suggest that the above analyzed \halpha~jets are extremely similar to spicules.

\subsection{\halpha~jets in the roots of coronal plumes}\label{subsec:plumes}
\begin{figure*}[!ht]
\centering
\includegraphics[trim=0cm 0cm 0cm 0cm,clip,width=\textwidth]{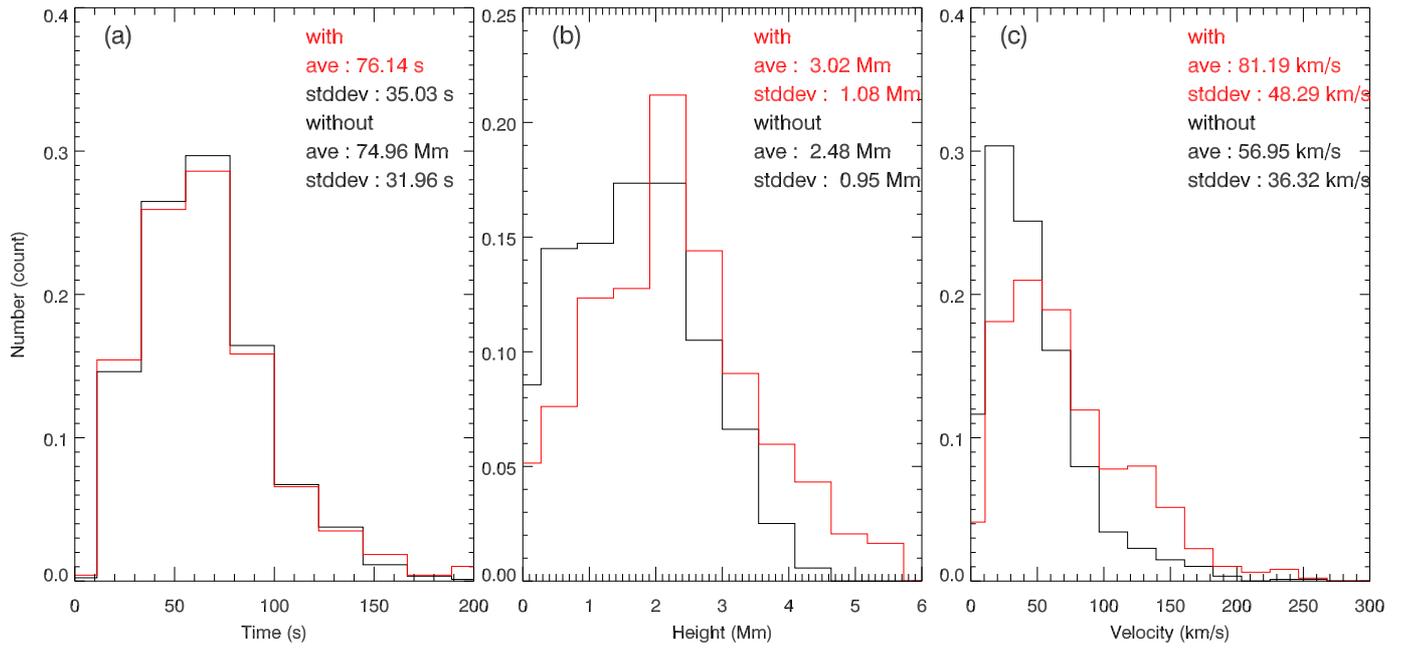}
\caption{The normalized distributions of the properties of (a) lifetimes, (b) heights and (c) velocities of the \halpha~jets identified and tracked in the regions of R1 - R5 (\textit{red lines}) and the rest regions (\textit{black lines}). The values following to``ave" and ``stddev" are the average values and standard deviations of the corresponding parameters, respectively.}
\label{figresult2}
\end{figure*}

In Fig.~\ref{figfov}, we see that plumes (plume-like features) are rooted in the regions of R1--R5,
while they are hardly seen in other regions as determined from the AIA 171\,\AA~images.
We also see coronal bright points are clearly shown in R1--R5.
In the \halpha\ images, we find 373 \halpha~jets rooted in R1--R5 and 947 rooted in the rest regions.
In Fig. \ref{figresult2}, we compare the statistics of \halpha~jets rooted in two kinds of regions.
The normalized distributions of the lifetimes in \halpha\ jets in these two regions are almost identical.
The significant differences are shown in the  distributions of heights and speeds.
Despite the distributions of each parameter from two kinds of regions are largely overlapped,
the distributions of heights and speeds from regions rich in coronal plumes (R1--R5) bias toward higher and faster wings than
those from the regions poor in coronal plumes.
Specifically, \halpha~jets in R1--R5 have height averaging at 3.01$\pm$1.07\,Mm, and speed averaging at 81.2$\pm$48.3\,\kms.
In contrast, the average height and speed of the \halpha\ jets rooted in other regions poor in coronal plumes are 2.48$\pm$0.95\,Mm and 56.9$\pm$36.3\,\kms, respectively.

\par
Our study in \paperi\ shows that transition region network jets in the footpoints of coronal plumes also tend to be more dynamic (higher and faster) than those in other regions. Together with the consistent results found here in \halpha\ jets, we might have two hypotheses: a cascade effect might exist in these phenomena that dynamic \halpha\ jets power network jets and then dynamic network jets power coronal plumes; alternatively, high speed \halpha\ jets, network jets and coronal plumes are three different counterparts of the same physical processes (e.g., compression of magnetic elements~\citep{2016ApJ...818..203W}).
A phenomenon consisting of multi-thermal structures is quite common in many coronal dynamics with more rapid evolution, such as coronal jets\,\citep[e.g.,][]{2015Natur.523..437S,2017ApJ...851...67S,2018ApJ...854...80H,2020ApJ...897..113H}, flaring events\,\citep[e.g.,][]{2014A&A...566A.148H,2018ApJ...853L..26H,2020MNRAS.498L.104W} and coronal bright points\,\citep[e.g.,][]{2012A&A...545A..67M,2018ApJ...869..175H},
apart from the relatively static events like coronal plumes.

\subsection{\halpha~jets and PDs in coronal plumes}\label{subsec:pds}
\begin{figure*}[!ht]
\centering
\includegraphics[trim=0cm 0cm 0cm 0cm,clip,width=\textwidth]{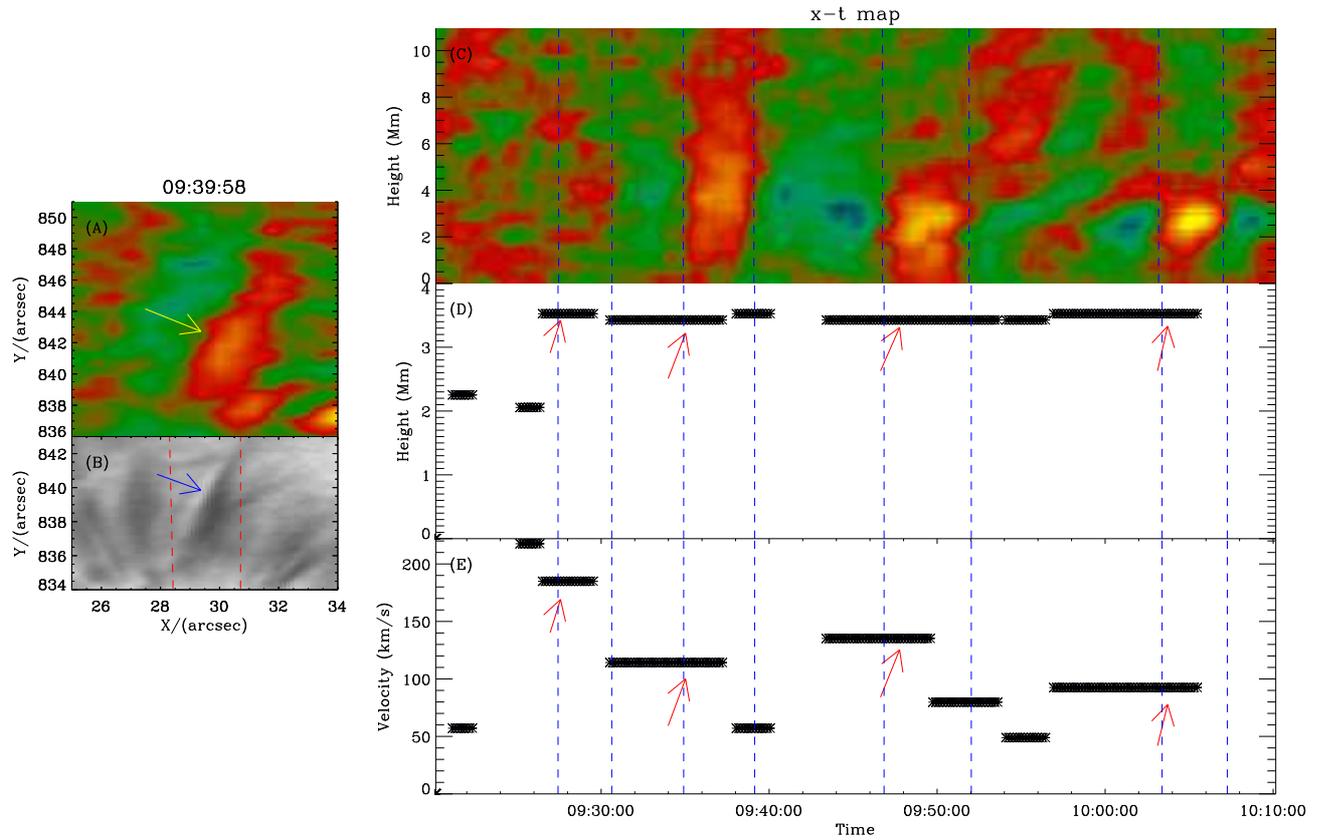}
\caption{A case analysis of the relationship between propagating disturbances (PDs) and \halpha~jet.
Snapshot showing the connection between \halpha~jets in the NVST
\halpha~-0.6\,\AA\ (B) and PDs in the filtered AIA 171\,\AA\ (A).The yellow arrow outlines
the structures of PDs determined from panel (A), while
the blue arrow represents the associated \halpha~jets from panel (B).
Time-slice images of the PD is shown in Panel (C). The blue lines mark the locations of PDs. The panel (D) $\&$ (E) show that the
the maximum height and ascending velocity of the~\halpha~jets which exist in the
footpoint of the given PD. Those denoted by the red arrows are the ones presented prior to the PD. }
\label{pd_1}
\end{figure*}

\begin{figure*}[!ht]
\centering
\includegraphics[trim=0cm 0cm 0cm 0cm,clip,width=\textwidth]{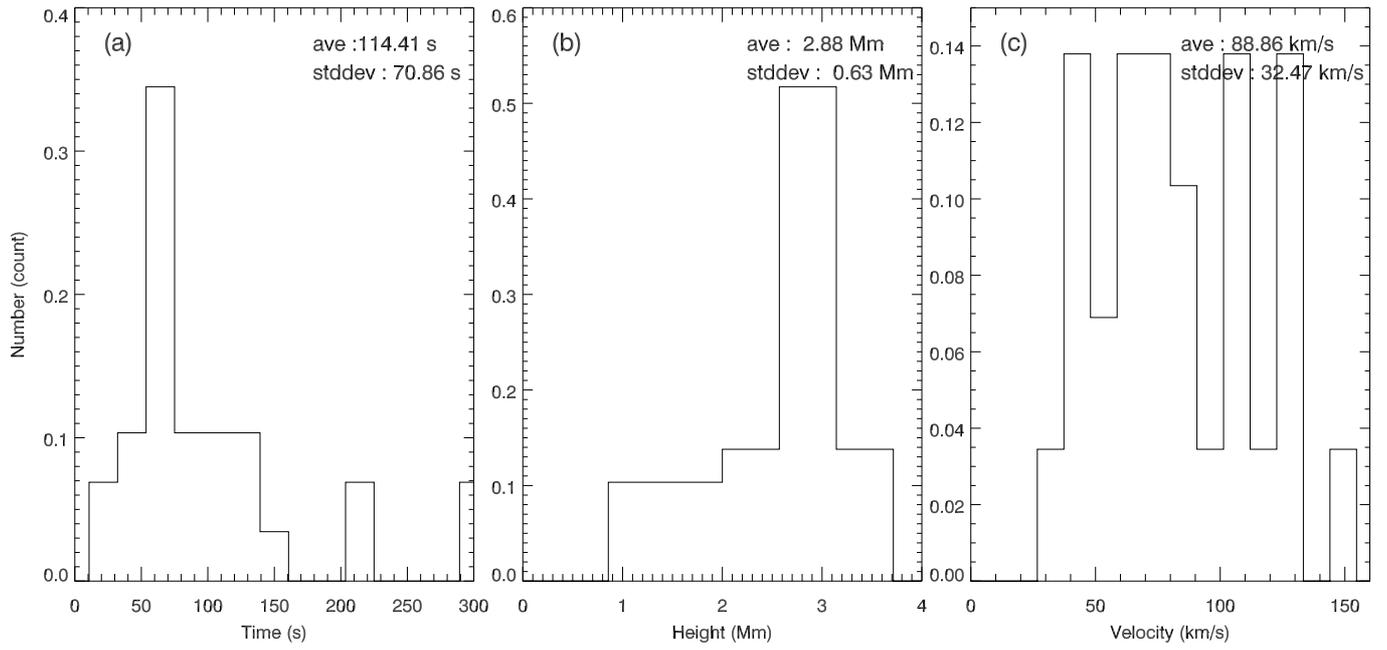}
\caption{The normalized distributions of the lifetimes (a), heights (b) and velocities (c) of the \halpha~jets which have
PDs response.}
\label{pd_sta}
\end{figure*}

In order to trace PDs, we apply the simple bandpass filter on the AIA 171\,\AA~
images. We filter out the signals with the period shorter than 10\,minutes and
longer than 30\,minutes. The detailed method is described in~\citet{2015ApJ...809L..17J},
and the results are shown in Fig.~\ref{figfov}(c).
In the associated animation of Fig.\,\ref{figfov}, PDs in R1--R5 can be clearly seen.
Since artificial periodicity near the cutoff periods might be introduced artificially\,\citep{2016ApJ...825..110A,2020A&A...634A..63K},
we choose manually and analyze further those PDs also showing discernible signals in the original images.
We further investigate in detail the connection between PDs and the \halpha\ jets in their footpoint regions, and an example is given in Fig.~\ref{pd_1}.
From Figs.~\ref{pd_1}(A)$\&$(B) and the associated animations,
we find that the footpoints of \halpha~jets and PDs root in the same region, and the directions of their extensions are consistent with each other.
For this reason, we believe that some \halpha\ jets and PDs might share the same magnetic flux tubes.
Figures~\ref{pd_1}(C) shows an time-distant diagram along the PDs (see yellow arrow in Fig.\,\ref{pd_1}(A)).The blue-dashed lines mark the locations of PDs.
Figure\,\ref{pd_1}(D)$\&$(E) display the maximum height and ascending speed of
the \halpha~jets along the same path as the PDs observed at different time.
Combined the panels (C), (D) and (E), we find
that there is always a \halpha~jet appearing before presence of a PD and also having the same orientation as that of the PD
(denoted by red arrows in panel (D) / (E)).
We also notice that many \halpha\ jets are rooted in the footpoints of PDs but not all of them are followed by PDs.
So there is an obvious question what kind of \halpha~jets are associated with PDs.
In Figs.\,\ref{pd_1}(D)$\&$(E), for the same series of PDs we can see that the \halpha~jets appear to be higher and/or faster before the presence of PDs.
For the series of PDs shown in Fig.\,\ref{pd_1}, the \halpha~jets appearing just before PDs have an average height of 3.5\,Mm and an average ascending velocity of 91.2\,\kms, in contrast to 3.0\,Mm and 73.9\,\kms\ for those appear at the other times.
This seems like that the~\halpha~jets corresponding to the PDs are more dynamic than those not associated with PDs.
To confirm this, we investigate all PDs identified in R1--R5, which include six series of them in total.
In these PDs, we trace 29 associated \halpha\ jets and show the histograms of their lifetimes, heights and speeds in Fig.\,\ref{pd_sta}.
Although the average height and speed of these \halpha\ jets do not have much difference from those rooted with coronal plumes, we can see that the almost all (except one) \halpha\ jets followed by PDs have speeds of about or more than 50\,\kms.
This implies that only those chromospheric jets with speeds $\gtrsim$50\,\kms\ can proceed PDs.
This conclusion, however, requires further observational, theoretical and/or numerical investigations.
The average lifetime of these \halpha\ jets is significantly larger, but this is mainly due to the small samples as there are four having lifetimes larger than 200\,s.

\par
We further select a cluster of \halpha~jets
existed in the vicinity of PDs (see between two red-dashed lines of Fig.~\ref{pd_1}(B)) and analyze their dynamics when the PDs appear.
We compare the average height and speed of the \halpha\ jets when PDs are shown and those when not shown.
We find that heights and speeds of the groups of \halpha~jets with PDs are greater than those of \halpha~jets not associated with PDs.
The \halpha\ jets associated with PDs have heights averaging at 3.47$\pm$0.06\,Mm, and speeds averaging at 91.2$\pm$32.8\,\kms. In contrast, the average height and speed of the \halpha\ jets not associated with PDs are 3.06$\pm$0.67Mm and 73.4$\pm$35.6\,\kms, respectively.
Since the cross-section of a PD is much wider than a single \halpha~jet, it is difficult to associate a particular jet to the PD. Therefore, it remains possible that PDs are proceeded by a set of \halpha\ jets, rather than a single one.
\subsection{The relation between~\halpha~jets and a coronal jet}\label{subsec:coronal jets}
\begin{figure*}[!ht]
\centering
\includegraphics[trim=0cm 0cm 0cm 0cm,clip,width=\textwidth]{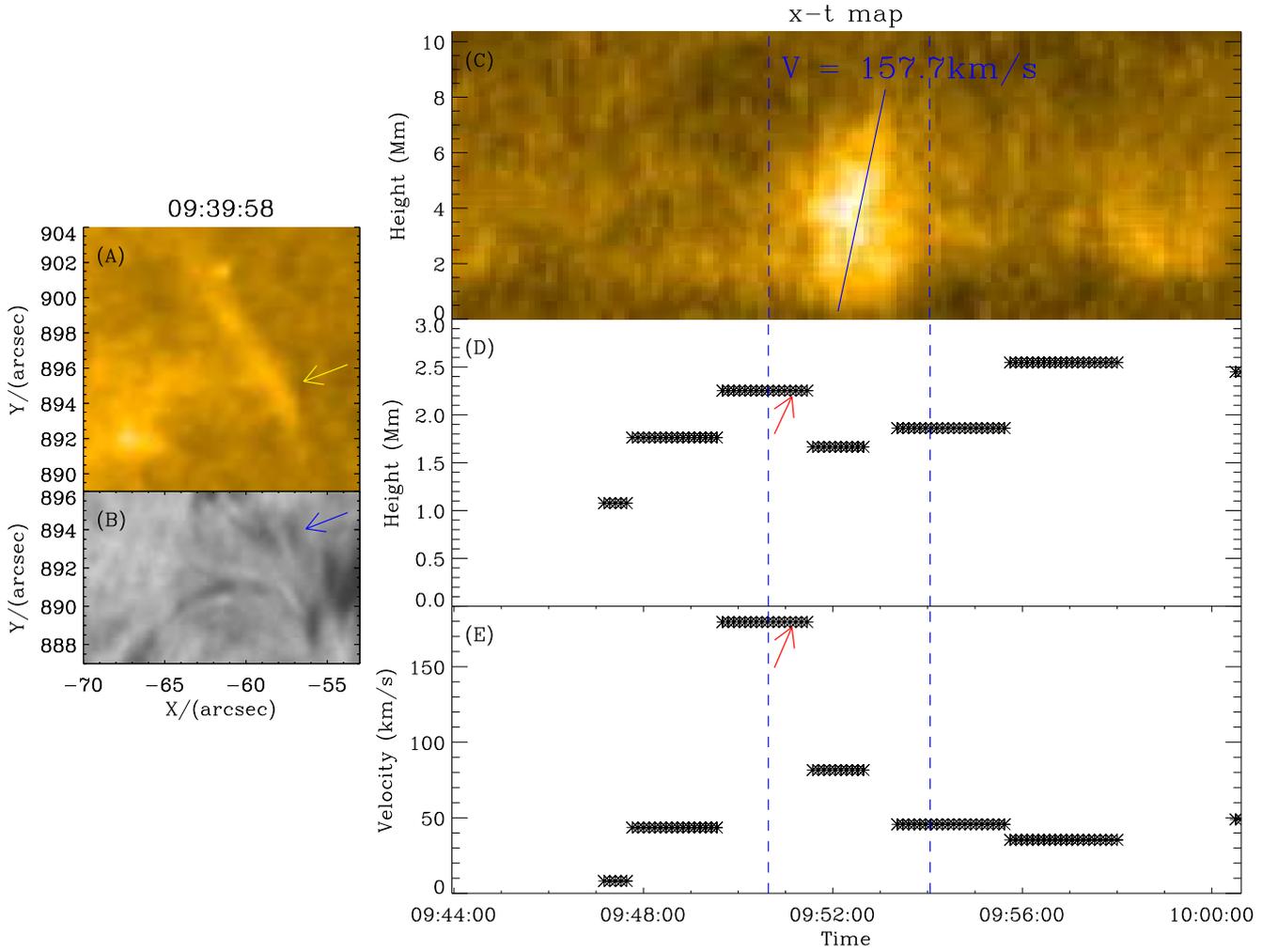}
\caption{A case analysis of the relationship between coronal jet and \halpha~jet.
Snapshot showing the connection between \halpha~jets in the NVST
\halpha~-0.6\,\AA\ (B) and coronal jet in the AIA 171\,\AA\ (A).The yellow arrow outlines
the structures of coronal jet determined from panel (A), while
the blue arrow represents the \halpha~jets from panel (B).
Time-slice images of the coronal jet is shown in Panel (C), the velocity is 158\,\kms.
The panel (D) $\&$ (E) show that the
the maximum height and ascending velocity of the~\halpha~jets which exist in the
footpoint of the given coronal jet (seen those denoted by the red arrows).
{\bf The blue dashed lines indicate the location of the coronal jet.}}
\label{coronaljet_1}
\end{figure*}

\citet{1998ApJ...509..461W} suggested that macrospicules seen in \halpha~jets and \heii~304 jets are also direct manifestations of magnetic reconnection.
Given that magnetic reconnection can trigger coronal jets, we investigate whether \halpha\ jets have any connection with a faint coronal jet or not.
In Fig.~\ref{coronaljet_1} and the associated animation, we show the evolution of the coronal jet observed in AIA 171\,\AA\ and the \halpha\ jets rooted in the same region.
From Fig.~\ref{coronaljet_1}(A), we can see that a coronal jet exists on the side of the footpoint of a coronal loop.
In the animation, we clearly see that the loop expands from the bright base and form a blowout coronal jet at 09:50:28\,UT, which can reach a height of about 9 Mm.
In Fig.~\ref{coronaljet_1}(C), the blue-dashed lines mark the location of the coronal jet,
the speed of the coronal jet is about 158\,\kms.
We also find that a \halpha~jet blows up from the footpoints of the loops at 09:48:30\,UT, just before initiation of the coronal jet (see Fig.~\ref{coronaljet_1}(B)).
 The development of the \halpha~jet is well consistent with that
of the coronal jet, including their spatial extensions and directions.
This \halpha~jet has a height of about 2.3\,Mm and a rising speed of about 180\,\kms.
The speed of the \halpha\ jet is similar with that of the coronal jet. Therefore,
we suggest that the coronal jets might be seen as the extension of part of \halpha~jets.
 Unfortunately, this coronal jet is the only case found in the present observations.
Further study in statistics is required to confirm our conclusion.

\section{Discussion}\label{sect:discussion}

In \paperi, we studied the relation between transition region network jets and coronal plumes.
In the present work, we focus on chromospheric jet-like features observed in \halpha\,--0.6\,\AA, namely \halpha\ jets. In agreement with \paperi, we found that the network regions where the coronal counterparts are active with coronal bright points and plumes tend to produce \halpha\ jets with more dynamic characteristics (i.e., larger heights and faster speeds) than those from the other network regions. Given that \halpha~jets, network jets and coronal plumes may be associated with magnetic reconnection, it is possible that the energy produced by magnetic activities could transfer from fine structures in the low solar atmosphere to those in the corona.

\par
In this paper, we find a few examples
that \halpha~jets have a one-to-one correspondence with PDs in spatial and temporal.
The \halpha~jets associated with PDs are usually faster and higher than the others.
What's more, \citet{1998SoPh..178...55W} found that \heii~304\,\AA~jets are always associated with \halpha~jets, but not all the \halpha~jets are associated with \heii~304\,\AA~jets.
We find that the faster and higher \halpha~jets have a certain relationship
with coronal jets. Perhaps, only parts of \halpha~jets which are more energetic can reach the low
corona and also be seen in \heii~304\,\AA.

\par
Why the \halpha\ jets are higher and faster at the roots of coronal plumes and jets? We will discuss the following possible mechanisms.

\par
 The first interpretation is associated with the generation of transverse wave induced KHi in \halpha~jets.
 It has been proved that chromospheric jets usually accompany transverse motions
 with amplitudes up to $10-25$\,\kms\,\citep{2007Sci...318.1574D}, which are generally interpreted as kink waves ~\citep{2009ApJ...705L.217H}.
 Such waves are possible to induce the KHi due to velocity shear between magnetic structures and background medium~\citep[e.g.,][etc.]{2008ApJ...687L.115T,2018ApJ...856...44A}.~\citet{2018ApJ...856...44A} considered the KHi induced by kink motions and attributed some observational properties of spicules to the development of KHi eddies.
 The small spatial structures induced by the KHi can help the dissipation of wave energy,
 and thus playing an important role in an understanding temperature increase of magnetic structures~\citep[e.g.,][]{2019ApJ...870...55G,2019A&A...623A..53K}. In addition,
 \citet{1980PhFl...23..939L} pointed out that if a longitudinal flow exceeds a criterion, the magnetized plasma is unstable.
 With the IRIS observations, \citet{2018NatSR...8.8136L} found that the KHi could develop due to the strong shear (more than 204\,\kms) between two blowout jets.
  In the case of kink waves with longitudinal flows,
 the criterion for developing the KHi has been discussed in many literatures,
 \citep[e.g.,][]{2001A&A...368.1083A,2016Ap&SS.361...51Z,2018NatSR...8.8136L}.
The threshold was derived by \citet[][]{2001A&A...368.1083A} from eigen-mode  analysis in a cylindrical model.
It reads
$U = V_{\rm Ai}(1+B_{\rm e}/B_{\rm i}\sqrt{\rho_{\rm i}/\rho_{\rm e}})$,
where $U$ ($V_{\rm Ai}$) represents the flow (Alfv\'en) speed inside a jet,
$B_{\rm i}$($\rho_{\rm i}$) and $B_e$($\rho_{\rm e}$) are the magnetic field strength (density) in a jet and the ambient region, respectively.
In spicules,
the density and magnetic field of a chromospheric jet is about 3 $\times 10^{10} {\rm cm}^{-3}$ and 40\,G, respectively.
Assume those for the ambient regions are 1 $\times 10^{9}{\rm cm}^{-3}$ and 10\,G
\citep[see][and references therein]{2012SSRv..169..181T,2000SoPh..196...79S,2010ApJ...708.1579C}.
Generally, the Alfv\'en speed in the choromosphere is 40\,\kms \citep{2007Sci...318.1574D},
indicating that the KHi can be induced in spicules when the upflow speed exceeds 98\,\kms.
Furthermore,~\citet{2016Ap&SS.361...51Z} studied the propagating of kink waves in EUV jets in vertical magnetic tubes.
 They found that kink waves could become unstable when flow velocity exceeds 112\,\kms.
 In Fig.~\ref{figresult2}, we can see that  28.4\% \halpha~jets have higher speed (more than 110\,\kms) in R1--R5, while only 8.3\% \halpha~jets with such high speed in other regions.
 This means that in aforementioned observations,
 the KHi is possible to develop in faster \halpha~jets in R1--R5.
 Due to the development of the KHi,
 the faster \halpha~jets are possible to dissipate energy into higher atmospheres.
Moreover,~\citet{2013NewA...25...89C} proposed that the KHi might be a mechanism for the formation of magnetic reconnection.
This means that the \halpha~jets with KHi might be able to heat the corona.
However, more detailed simulations should be considered in future studies.

\par
Another possible interpretation is the interaction strength among small-scaled magnetic elements in the network region.
For R1-R5, there are dominant network fields and some small-scale opposite-polarity fields in Fig.~\ref{figfov}(b).~\citet{2016ApJ...818..203W} proposed that the convergence of the base flux could trigger coronal plumes. The interaction of magnetic flux concentrations with constantly weak fields could result in strong magnetic tension. The strong magnetic tension which is amplified and transported upward through ambipolar diffusion could produce numerical faster spicules (more than 100\,\kms) in 2.5D simulation~\citep{2017Sci...356.1269M}. In addition, same as the 2.5D numerical simulation,~\citet{2011A&A...535A..95D} also showed that the stronger magnetic field could produce higher up-flows velocities. These simulations mean that there are a great possibility of rapid jets in network field with weak opposite magnetic elements,
 and these jets might have coronal connection.
 Using high-spatial-resolution and high-time-cadence observations of the Goode Solar Telescope~\citep[GST,][]{2010AN....331..620G,2010AN....331..636C}, ~\citet{2019Sci...366..890S} provided observations from the photosphere to the chromosphere and the corona and supported that these chromospheric jets ($\sim$50\,\kms) are driven by the interaction between strong network fields with small-scale opposite-polarity fields, of which have a good coronal connection. By performed order-of-magnitude estimation of energies, the authors estimated that these spicules supply enough energy into the corona. These speeds of jets are smaller than that in our results because the FOV in ~\citet{2019Sci...366..890S} is close to the center of disk. With respect to the relation between the chromospheric jets and coronal activities,
 the mixed-polarity magnetic fields could result in shuffling and further braiding of small-scale magnetic field lines, which produce magnetic reconnection at braiding boundaries and power the corona. By the same,
  ~\citet{2018ApJS..237....7L} proposed that the dissipation of electric currents due to the effect of small-scale weak magnetic activities (included spicules) could effectively heat corona at low heights. In addition,~\citet{2002ApJ...576..533P} showed that the formation and dissipation of current sheets along these separatrices resulting from highly fragmented photospheric magnetic configurations could be an important contribution to coronal heating. This means that the activities of mixed-polarity on small scales might have enough power to coronal activities.
 In Fig.~\ref{figfov}(d), we see that \halpha~jets locate near the dominant network fields, but the lower resolution of magnetic fields measurements affect the detection of weak magnetic elements at the footpoints of \halpha~jets.
 Further studies using high spatial-temporal resolution observations of the photosphere (such as the Goode Solar Telescope (GST), the forthcoming Daniel K. Inouye Solar Telescope (DKIST) and the planned Chinese Advanced Solar Observatories--Ground-based (ASO-G)) should shed more light on this topic.

\par
Our analyses here have shown that \halpha\ jets with coronal responses tend to be higher and faster.
Whether there are critical height and speed for a \halpha\ jet having coronal response should be investigated both theoretically and observationally.
As discussed above, it may have different answers based on different mentioned mechanisms.
Based on the present observations, we cannot find evidence of KHi due to the spatial and temporal resolutions. HMI magnetograms show clearly that a number of small-scale magnetic elements appear and disappear in the footpoints of these \halpha\ jets (see the animation associated with Fig.\,\ref{figfov}). Although the observed region is pretty much closed to the limb, the observations still give a hint of a degree of magnetic dynamics. Therefore, with the present observations, we are more favorable with the scenario of that more powerful interaction between small-scale magnetic elements tends to generate more dynamic phenomena. However, the KHi scenario cannot be ruled out within the present circumstance.
In the near future, we will study in-depth how much initial energy of chromospheric jets is required to have coronal responses based on the above-mentioned mechanism.

\section{Conclusions}\label{sect:conclusions}
In this paper, we analyzed observations taken by the 1-m New Vacuum  Solar Telescope at a passband \halpha\,--0.6\,\AA. Thanks to the higher spatial and temporal resolution of the data,
we were able to trace the evolutions of \halpha~jets over time.
Using an upgraded automated detection algorithm, we carried out a statistical study of \halpha~jets.
Using the algorithm, we identified and traced 1\,320 \halpha~jets in the data set.
We found that the average lifetime, height and
ascending speed of the \halpha~jets are 75.38\,s, 2.67\,Mm, 65.60\,\kms, respectively. These characteristics of the events agree well with those of \typeii~spicules.
In addition, we found that the \halpha~jets occur in network regions where magnetic concentrations are present.
We believe that the \halpha~jets are corresponding to spicules or the on-disk counterparts of spicules.

\par
We found that \halpha\ jets rooted in the footpoint regions of coronal plumes have an average lifetime of 76.1$\pm$35.0\,s, an average height of 3.01$\pm$1.07\,Mm and an average ascending velocity
of 81.2$\pm$48.3\,\kms.
While the \halpha~jets in the rest regions are having an average lifetime of 75.0$\pm$32.0\,s, an average height of 2.48$\pm$0.95\,Mm and an average ascending velocity of 56.9$\pm$36.3\,\kms, these results show that the \halpha~jets associated with coronal plumes are on average higher and more dynamic than those not with coronal plumes.
Can these jets possibly provide energy to sustain a coronal plume?
In corona, radiation cooling time is expressed as $\tau_{rad}(n_0,T_0)=\frac{9}{5}\frac{k_BT^{5/3}}{n\Lambda_0}$ in the corona\,\citep{2008ApJ...686L.127A}, taking in typical parameters of corona and plumes, where the radiative loss rate $\Lambda_0\sim10^{-17.73}ergs/cm^{3}$,
temperature T$\sim$1.0\,MK and density n$\sim10^{9}cm^{-3}$, then we found the radiation cooling time scale is in the order of 1000\,s. While the \halpha~jets in the footpoint of a plume studied here present repeatedly in a time scale of about 80\,s, much less than the radiation cooling time in the corona,
it suggests that \halpha~jets may provide energy continuously to sustain coronal plumes.

\par
To determine exactly what type of \halpha\ jets might link to coronal activities, we further investigated the connection between \halpha\ jets and the PDs in plumes.
We found that there are always \halpha\ jets appearing before the initiation of PDs and extending in the same directions of the PDs.
Almost all these jets (28 out of 29) have speeds $\gtrsim$50\,\kms, thus we suggest that only those \halpha\ jets exceeding a certain velocity can be precursors of PDs.

\par
A fade coronal jet with a speed of about 160\,\kms\ was also observed.
We found that this coronal jet was accompanied by a \halpha~jet that has a speed comparable to that of the coronal jet,
suggesting that the coronal jet might be the extension of the part of \halpha~jets.

\par
In agreement with \paperi, present studies confirmed that small-scaled jet-like features in the solar lower atmosphere can directly connect to coronal activities.
Given that most \halpha\ jets observed here are likely spicules in the chromosphere, we suggest that \halpha~jets, if dynamic to a certain degree, can sufficiently power the corona.
Combined the results shown in \paperi, we suggest that spicule-like features in the chromosphere, transition region network jets and ray-like features in the corona are coherent phenomena, and they are important tunnels for cycling energy and mass in the solar atmosphere.

\par
{\it Acknowledgments:}
We are grateful to the anonymous reviewer for the constructive comments and suggestions.
L.X. and Y.Q. are supported by National Natural Science Foundation of China (NSFC) under contract No. 41974201.
Z.H. thanks financial supports from NSFC under contract No. U1831112 and the Young Scholar Program of Shandong University, Weihai (2017WHWLJH07).
H.F. is supported by NSFC contract No. U1931105.
W.L. is supported by NSFC contract No. U1931122.
M.S. is supported by NSFC contract No. 41627806.
We would like to thank the NVST operation team for their hospitality during the observing campaign and preparation of the data.
The data used are also courtesy of NASA/SDO, the AIA and HMI teams and JSOC.

\bibliographystyle{aa}
\bibliography{chrom_plume_connection}

\end{document}